# Thorium doped strontium fluoride crystal: a unique candidate for solid nuclear optical clock material

Qiaorui Gong,[1,2] Shanming Li,[1] Shulong Zhang,[1] Siliang Tao,[1] Guoliang Deng,[1,2] Peixiong Zhang,[3,*] Chengchun Zhao,[1,2,†] Yin Hang,[1,2,‡] Shining Zhu,[4] and Longsheng Ma[5]

[1]Research Center of Laser Crystal, Shanghai Institute of Optics and Fine Mechanics, Chinese Academy of Sciences, Shanghai 201800, China

[2]Center of Materials Science and Optoelectronics Engineering, University of Chinese Academy of Sciences, Beijing 100049, China

[3]Department of Optoelectronic Engineering, Jinan University, Guangzhou, Guangdong 510632, China

[4]National Laboratory of Solid-State Microstructures, School of Physics, Collaborative Innovation Center of Advanced Microstructures, Nanjing University, Nanjing, Jiangsu 210093, China.

[5]State Key Laboratory of Precision Spectroscopy, East China Normal University, Shanghai 200062, China

*pxzhang@jnu.edu.cn

†zhaocc205@siom.ac.cn

‡yhang@siom.ac.cn

We report a candidate with unique advantages in the cultivation of solid-state nuclear clock material, Th:SrF$_2$ crystal. It not only has a segregation coefficient close to 1, which can achieve highly efficient and uniform doping of Th, but also ensures a high transmittance (~69% at 150 nm) while achieving extremely high doping concentration ($^{232}$Th>6×10$^{20}$ cm$^{-3}$). In addition, SrF$_2$ crystal will not be irradiated-colored under strong α radiation like CaF$_2$ crystal, Th:SrF$_2$ crystal is expected to fully unleash its high concentration doping characteristics while ensuring its transmission performance in nuclear transition band not be severely affected by $^{229}$Th radiation damage.

Thorium-229 ($^{229}$Th) and its anomalously low-energy isomer state $^{229m}$Th points to a new type of clock based on nuclear transition frequency - nuclear optical clock [1]. After nearly 20 years of development, scientists have made milestone progresses and are gradually approaching this goal [2-25]. Among them, the solid-state scheme of nuclear clocks has demonstrated superiority in nuclear spectroscopy research. Relying on $^{229}$Th doped crystals, scientists have achieved direct laser excitation and deexcitation signal detection of $^{229m}$Th [3-5]. This undoubtedly provides a strong shot for the development of nuclear clocks.

The cultivation of $^{229}$Th doped crystals with excellent comprehensive performance is crucial for the development of solid-state nuclear clocks. However, due to the α-decay of $^{233}$U ($T_{1/2}$=1.6×10$^5$ years) is currently the only means to obtain usable quantities of $^{229}$Th, which makes its reserves extremely small and price expensive. Although, $^{233}$U was produced in large quantities (1556 kg) with the idea to fuel nuclear power reactors at the Savannah River Site (USA) in the late 1950s [26], only 45 g of $^{229}$Th should have been produced since then and very small amounts (~mg) are available due to its labor-intensive chemical separation and limit by chemical impurities [27]. Moreover, $^{229}$Th has the extreme toxicity and strong radioactivity, which requires strict radiochemical experimental conditions for conducting related experiments. All these factors make the efficient cultivation of $^{229}$Th doped crystals difficult. Most studies on Th doped crystals first use natural thorium, $^{232}$Th, which has the same electronic and chemical properties as $^{229}$Th, as well as high reserves, moderate toxicity, low radioactivity, and stronger operability, as a dopant to evaluate the feasibility and potential of a material as a candidate for solid-state nuclear clocks material. Further research on the cultivation and properties of corresponding $^{229}$Th doped crystal will only be carried out for the matrix crystal that have been identified as having good potential.

The segregation coefficient of the candidate crystals for Th ions can be used to evaluate their doping ability, a low segregation coefficient leads to weak doping, high loss and baddish uniform of Th ions. Among the publicly available Th-doped crystals [27-47], Th:CaF$_2$ crystals have relatively stronger doping ability, it has a segregation coefficient of about 0.3 and can achieve doping levels of >10$^{20}$ cm$^{-3}$ [27,29,31]. However, current attempts at $^{229}$Th:CaF$_2$ crystals are limited by the limited availability of $^{229}$Th and factors such as crystal coloration and deterioration of VUV transmittance caused by $^{229}$Th radioactive irradiation. The doping concentration of $^{229}$Th is still limited to the level of ~10$^{18}$ cm$^{-3}$ [27,31], and its high doping and high transmittance characteristics have not been truly utilized. The latest literature suggests that annealing in a fluoride atmosphere can alleviate this problem to some extent [32]. Overall, the doping efficiency, doping loss, and doping uniformity of these existing crystals (include Th:CaF$_2$ crystal) are still not ideal, and there are also challenges in balancing their doping concentration and transmittance. All these limits are unfavorable for the already extremely scarce $^{229}$Th raw material and the development $^{229}$Th doped crystals. If a matrix material can be found that can achieve highly efficient, low loss and uniform doping of $^{229}$Th, and has certain advantages in VUV transmittance, radiation damage resistance, and background luminescent noise, it will be beneficial for the research, development, application, and subsequent promotion of solid nuclear clock materials.

Here, we report a unique solid-state nuclear clock candidate material, Th:SrF$_2$ crystal, with significant advantages in Th doping, growth, and performance. The structural changes, defect properties, segregation characteristic, and VUV transmittance performances of Th:SrF$_2$ crystals were studied based on the theoretical calculations and experiments. The results shows that the Th:SrF$_2$ crystal has a segregation coefficient for Th close to 1, can ensure highly efficient, uniform and low loss doping of Th, which will greatly reduce the difficulty of cultivating solid-state nuclear clock materials. The cultivated $^{232}$Th:SrF$_2$ crystal exhibits high VUV transmittance (~69% at 149 nm) at a high doping level (>6×10$^{20}$ cm$^{-3}$), reflecting excellent tolerance for Th nuclei. Moreover, the $^{229}$Th:SrF$_2$ crystal is expected to fully unleash its high concentration doping characteristics while ensuring that its transmission performance in nuclear transition band is not severely affected by $^{229}$Th radiation damage. The discovery of this crystal will greatly promote the development of solid-state nuclear clock materials and it is expected to be one of the preferred materials for future solid-state nuclear clock research.

Prior to the disclosure of Th:SrF$_2$ crystals in this article, as reviewed earlier, Th:CaF$_2$ crystals were the most favorable candidate materials for achieving high concentration doping and effective utilization of Th ions. Its segregation coefficient (~0.3) is higher than other reported candidates [27,29]. However, this is still less than expected because the radius difference between Ca ions and Th ions is very small, and their electronegativities are also very close, it is reasonable to expected that Th:CaF$_2$ can achieve stronger doping ability. The only possible limitation is the impact of heterovalent doping. Our latest study on Th:CaF$_2$ shows that although the formation energy of Th containing composite defects is low when interstitial F serves as a charge compensation mechanism for heterovalent doping, the electrostatic repulsion between interstitial F ions and lattice F ions will cause significant local distortion and lattice expansion [29]. This is actually not conducive to the stability of the crystal, and also means that during the formation process of such defects, a large dynamic potential barrier will be encountered, which will limit the formation of the defects, thereby limiting the doping ability of Th in Th:CaF$_2$.

Based on the above considerations, we realize that SrF$_2$ with the same crystal structure as CaF$_2$ may have unique potential in the utilization of Th. Due to the smaller radius of Th ions compared to Sr ions, it will cause local lattice contraction when Th ions replace Sr ions into the lattice, which can compensate for the local lattice expansion caused by the interstitial F charge compensation mechanism. This will reduce the dynamic potential barrier during the formation process of the defects with favorable energy, making it easier for Th ions to be introduced into the lattice of SrF$_2$.

For the study of heterovalent doping, the previous work has provided a good inspiration. The charge compensation mechanism of heterovalent doped ion tends to locate at its first nearest neighbor position [3,37,48-50]. Here, we mainly considered the first nearest neighbor configurations of Th-related defects in Th:SrF$_2$ crystal. The Sr vacancies (V$_{Sr}^{2-}$) and interstitial F ions (F$_{in}^-$) located at the first nearest neighbor positions of Th are consider as charge compensation mechanisms of Th$_{Sr}^{2+}$. One Th$_{Sr}$-V$_{Sr}$ and two Th$_{Sr}$-2F$_{in}$ (180° and 90°) defect models are constructed in 2×2×2 supercells of SrF$_2$ crystal, respectively. **Figure 1** shows the theoretical models used for the calculations. Figure 1a shows the local coordination polyhedron model of Th$_{Sr}^{2+}$ without charge compensation. On this basis, Figure 1b shows the Th$_{Sr}$-V$_{Sr}$ defect model, consisting of a Th ion in substitution sites and a Sr vacancy. Figure 1c shows the Th$_{Sr}$-2F$_{in}$ (90°) defect model. The angle between the two interstitial F ions and Th ions is 90°, forming a right-angle configuration. Figure 1d shows the Th$_{Sr}$-2F$_{in}$ (180°) defect model. The angle between the two interstitial F ions and Th ions is 180°, forming a linear configuration. The calculation details can be seen in the Supplemental Material [51] (see also references [52-57] therein).

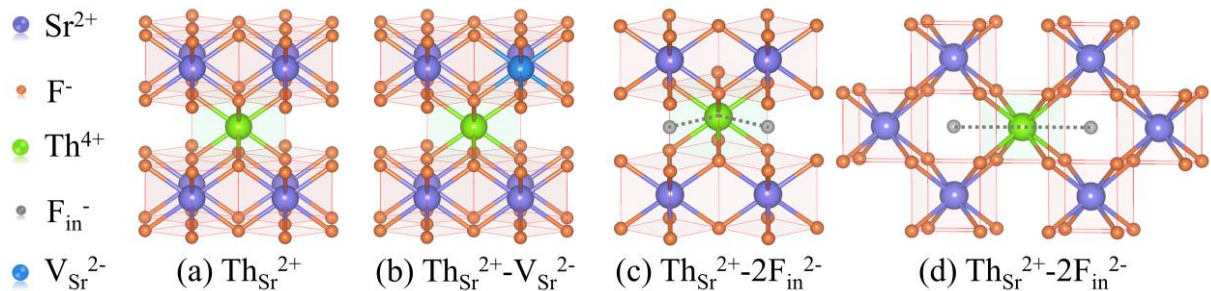

**Figure 1.** (a) The local coordination polyhedron model of Th$_{Sr}^{2+}$ without charge compensation; (b) the Th$_{Sr}$-V$_{Sr}$ defect model; (c) the Th$_{Sr}$-2F$_{in}$ (90°) defect model; (d) the Th$_{Sr}$-2F$_{in}$ (180°) defect model. (The purple, orange, grey, green and blue balls represent Sr atoms, F atoms in lattice sites, F atoms in interstitial sites, Th atoms and Sr vacancies, respectively. All the defect models are constructed in 2×2×2 supercells of SrF$_2$ single crystal.)

**Table 1** presents the calculation results of the structural changes and formation energies of the theoretical models. It should be noted that the construction of a single theoretical model has specific orientations, which leads to differences in its influence on cell parameters in different directions. However, there are often the same defect configurations with various orientations in real crystals, which will eliminate the orientation differences of a single theoretical model. Therefore, we considered the impact of orientation differences during model construction by taking the average value of the cell parameters a, b, and c of the theoretical models when calculating the change rates of the cell parameters (Δa). In terms of structure, by comparing the changes in cell parameters and cell volume between the Th:SrF$_2$ and pure SrF$_2$ models, it can be seen that all the three defect models result in an increase in lattice parameters and unit cell volumes. The Th$_{Sr}$-2F$_{in}$ (180°) defect model induces the least change in lattice parameters and volume (Δa=0.116% and ΔV=0.347%), followed by the Th$_{Sr}$-2F$_{in}$ (90°) defect model (Δa=0.261% and ΔV=0.783%), while the Th$_{Sr}$-V$_{Sr}$ defect model leads to relatively pronounced unit cell expansion (Δa=0.322% and ΔV=0.970%). In terms of the defect formation energy, the defect configurations using interstitial F as charge compensation mechanisms have lower formation energies, which means that, thermodynamically, it tends to prioritize the formation of such defects.

**Table 1.** The structural changes and defect formation energies of Th:SrF$_2$ and Th:CaF$_2$ crystal with the same defect configuration. (Δa and ΔV is the change rates of lattice parameters and unit cell volumes of Th:SrF$_2$ or Th:CaF$_2$ defect models relative to pure SrF$_2$ or CaF$_2$ model.)

| Models | Cell parameters a or a/b/c [Å] | Δa [%] | Cell volumes V [Å$^3$] | ΔV [%] | Formation energy [eV] | Reference |
|---|---|---|---|---|---|---|
| Pure SrF$_2$ | 5.85568 | - | 200.78488 | - | - | |
| Th$_{Sr}$-V$_{Sr}$ | 5.87536/5.87536/5.87291 | 0.322 | 202.73155 | 0.970 | 5.34 | This work |
| Th$_{Sr}$-2F$_{in}$ (90°) | 5.86795/5.86795/5.87690 | 0.261 | 202.35766 | 0.783 | -7.45 | |
| Th$_{Sr}$-2F$_{in}$ (180°) | 5.86584/5.86075/5.86075 | 0.116 | 201.48203 | 0.347 | -6.60 | |
| Th$_{Ca}$-V$_{Ca}$ | - | 0.610 | - | 1.839 | 5.66 | |
| Th$_{Ca}$-2F$_{in}$ (90°) | - | 0.696 | - | 2.100 | -6.68 | [29] |
| Th$_{Ca}$-2F$_{in}$ (180°) | - | 0.527 | - | 1.588 | -5.87 | |

Here we want to make a comparison. We used to conducted similar studies on Th:CaF$_2$ crystals, constructing the same types of defect models containing Th within the same sized supercells, and studying their impact on the structure of Th:CaF$_2$ crystals [29]. As listed in **Table 1**, the effect of the defect models on the cell parameters and cell volume of Th:SrF$_2$ crystals is significantly smaller than that on Th:CaF$_2$ crystals. For the defect configurations that use interstitial F as compensation mechanisms, the difference is particularly evident. It can be seen that compared to the Th$_{Ca}$-2F$_{in}$ (180°) defect model, the structural changes caused by the Th$_{Sr}$-2F$_{in}$ (180°) model are reduced by nearly 5 times. This confirms our discussion before, namely when Th ions with smaller radius replace Sr ions into the lattice, it will cause local lattice contraction, which can compensate for the local lattice expansion caused by the interstitial F charge compensation mechanism and then reduce the influence of composite defects on the crystal structures. This will reduce the dynamic potential barrier during defect formation and promote the formation of defect configurations with favorable energy, making it easier for Th ions to be introduced into the lattice of $SrF_2$.

When determining whether a defect is easy to form, it is necessary to consider both kinetic and thermodynamic factors comprehensively. This has been discussed in detail in our previous research and will not be further elaborated here [28-30]. To take all the dynamic and thermodynamic factors that has been discussed in this section into consideration, the conclusion can be drawn that the interstitial F mechanisms may be the primary charge compensation mechanisms for Th ions in the Th:$SrF_2$ crystal, and the $Th_{Sr}$-$2F_{in}$ (180°) defect model represents a superior defect configuration. This also implies that Th ions in the $SrF_2$ lattice tend to form a 10-coordination configuration with F ions, which will constraint nonradiative transition pathways of Th nuclei. The conclusion is similar to our previous research on Th:$CaF_2$ crystal [29], but the difference is that the related defect configuration in Th:$SrF_2$ crystal not only has a lower formation energy, but also has a very small impact on the crystal structure, that is, the difficulty of forming such defects in both thermodynamics and kinetics is very low. This indicates that Th:$SrF_2$ crystal may have stronger Th doping ability than Th:$CaF_2$ crystal and can achieve higher segregation coefficient and doping level of Th.

In this section, the electronic and optical properties of Th:$SrF_2$ crystal containing $Th_{Sr}$-$2F_{in}$ (180°) defect model are further investigated to evaluated the effects of Th doping on its VUV transmittance performance. In order to overcome the serious underestimation of crystal bandgap by traditional PBE functional and obtain a bandgap closer to its actual value, we adopted the PBE0 hybrid functional to calculate the electronic and optical properties of the crystals. **Figure 2** shows the band structure, sum density of states (sum DOS) and partial density of states (PDOS) of the pure $SrF_2$ crystal and the Th:$SrF_2$ crystal containing the $Th_{Sr}$-$2F_{in}$ (180°) defect model. As can be seen that the PBE0 band gap of pure $SrF_2$ crystal is 9.782eV, although it is still underestimated compared with the experimental value 11.25 eV [58], it is already quite close. The PBE0 band gap of Th:$SrF_2$ crystal is 8.941 eV, which is only slightly less than that of pure $SrF_2$ crystal. Considering the theoretical underestimation, the band gap of Th:$SrF_2$ crystal is obviously larger than the energy of Th isomer state. By comparing the sum DOS and PDOS of the pure $SrF_2$ crystal and the Th:$SrF_2$ crystal, it can be found that Th ion doping mainly introduces Th-6d defect energy levels near the bottom of the conduction band of Th:$SrF_2$ crystal, which is the main reason for the decrease of the band gap.

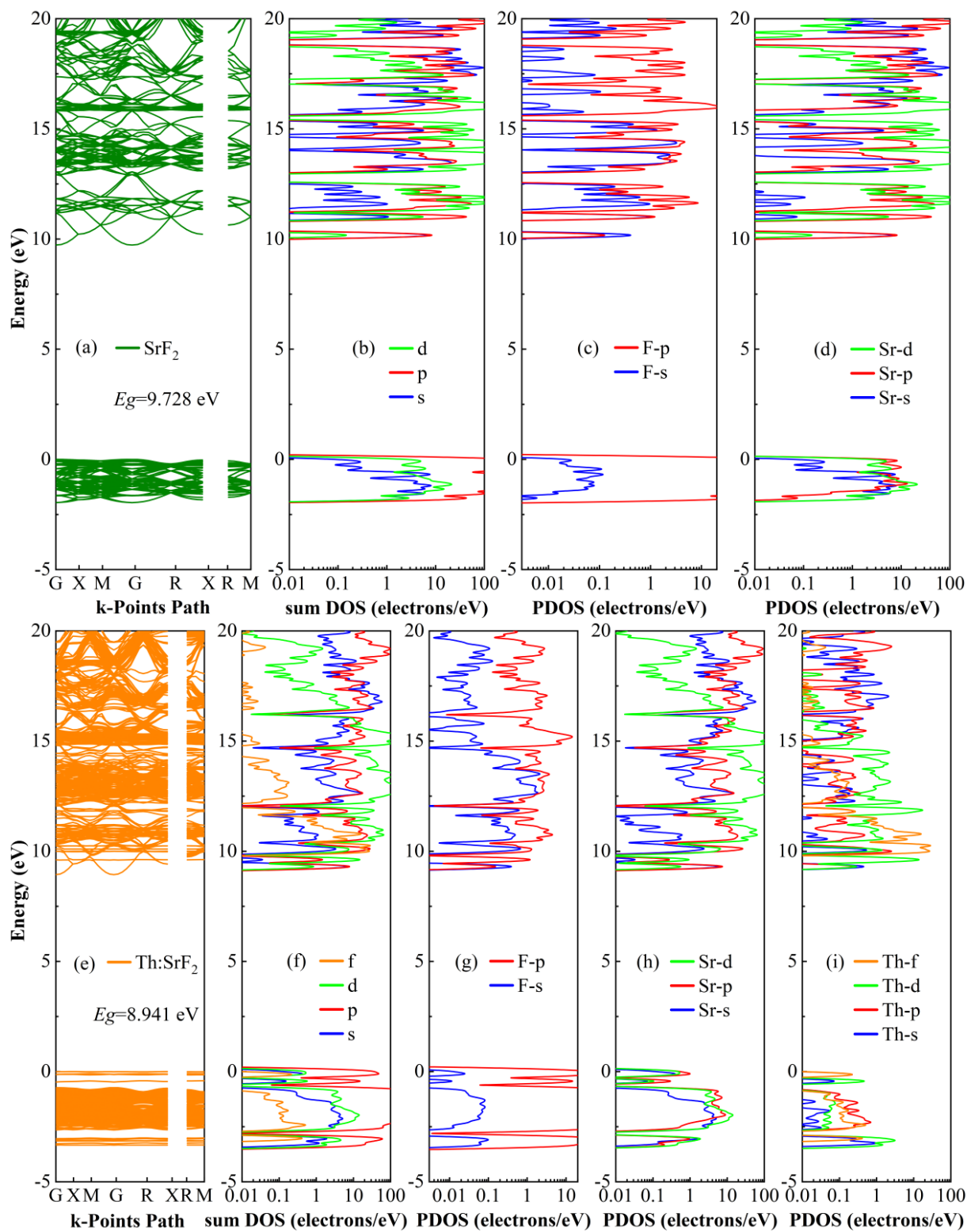

**Figure 2.** (a)-(d) the band structure, sum DOS and PDOS of pure SrF$_2$ crystal; (e)-(i) the band structure, sum DOS and PDOS of Th:SrF$_2$ crystal containing Th$_{Sr}$-2F$_{in}$ (180°) defect model.

In order to more clearly identify the orbital components of the defect energy levels, we further calculated the charge density distribution of its highest occupied molecule orbital (HOMO) and lowest unoccupied molecule orbital (LUMO) of pure SrF$_2$ crystal and the Th:SrF$_2$ crystal. The results are shown in **Figure 3**. Compared with the pure SrF$_2$ crystal, the HOMO of Th:SrF$_2$ crystal has the additional orbital components of F$_{in}$-2p belonging to interstitial F, while the LUMO of Th:SrF$_2$ crystal has the different orbital components, which are mainly Th-6d and F-3p rather than F-3s and Sr-5s. This further supports the conclusion that Th-doping introduces Th-6d defect energy level at the bottom of the conduction band. Overall, Th doping only slightly reduces the band gap, which will lead to the redshift of the UV cutoff edge, but since the band gap is still significantly larger than the energy of Th isomer state, we expect the effect of Th doping on the transmittance of SrF$_2$ crystal in the nuclear transition band to be limited.

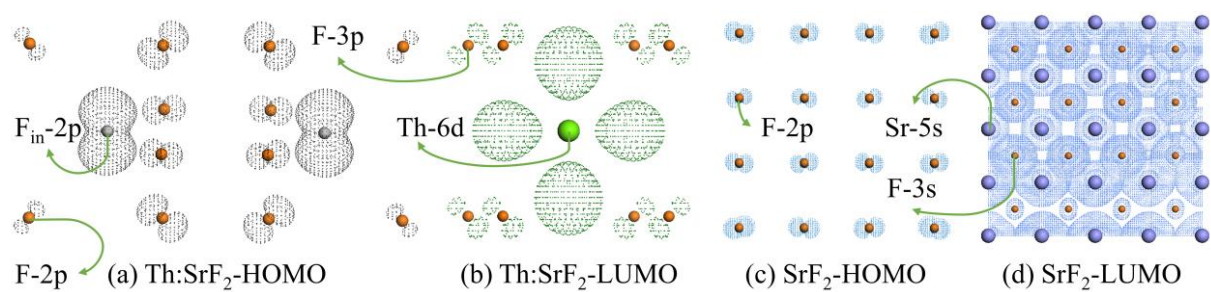

**Figure 3.** The charge distributions of HOMO and LUMO of the Th:SrF$_2$ (a,b) and pure SrF$_2$ crystals (c,d).

The theoretical results support that the Th:SrF$_2$ crystal has stronger Th doping ability than Th:CaF$_2$ crystal. Meanwhile, Th doping will not significantly affect the transmittance performance of SrF$_2$ crystal in the nuclear transition band. The Th:SrF$_2$ crystal is a highly promising candidate for solid-state nuclear clock material and is expected to achieve comprehensive performance superior to other candidates, especially in terms of doping.

Based on the theoretical research results, we directly grew high concentration doped $^{232}$Th:SrF$_2$ crystal by the Czochralski method and characterized its properties. The detailed description about growth and characterization can be found in the Supplemental Material [51]. The as-grown Th:SrF$_2$ crystal is very transparent and of good quality, and there was no obvious scattering centers. In addition, the remaining raw material in the crucible also spontaneously crystallized into transparent crystals after 24 h of slow cooling. This is completely different from other Th doped crystals we have grown in the past. There is no white powder residue or snowflake like polycrystalline at the bottom of the crucible, which is a characteristic that only systems with a segregation coefficient close to or greater than 1 and excellent crystallization performance possess. Therefore, we speculated that the segregation coefficient of Th in Th:SrF$_2$ crystal is close to or greater than 1. We took samples and tested the Th content inside the top and bottom of the crystal to further confirm this discovery.

**Table 2** lists the Th content (weight percentage) and corresponding doping concentrations of the samples obtained by ICP-AES (Inductively Coupled Plasma Atomic Emission Spectrometry) testing. The doping concentration at the top of the crystal is higher than that at the bottom of the crystal, and both the doping concentrations at the top and bottom are higher than the initial ratio of the raw materials, which is consistent with the characteristic that the segregation coefficient is greater than 1. By utilizing the top content and initial ratio, we can obtain that the segregation coefficient of Th in Th:SrF$_2$ crystals is approximately 1.14. Such excellent doping ability and such high doping level (>6×10$^{20}$ cm$^{-3}$) have never been achieved before. This means that the cultivating of $^{229}$Th:SrF$_2$ crystals can achieve highly efficient, uniform and low loss utilization of $^{229}$Th, which will greatly alleviate the extremely scarce problem of $^{229}$Th and promotes the development of solid-state nuclear clocks materials.

**Table 2.** Thorium content in initial ingredients and the as-grown Th:SrF$_2$ crystal. (Note: top and bottom means the sampling positions on the same crystal)

| Samples | Initial ingredients | Th:SrF$_2$ (top) | Th:SrF$_2$ (bottom) |
| --- | --- | --- | --- |
| Weight percentage of Th (wt. %) | 5.31 | 6.06 | 5.72 |
| Doping concentration of Th (cm$^{-3}$) | 6.10×10$^{20}$ | 6.66×10$^{20}$ | 6.29×10$^{20}$ |

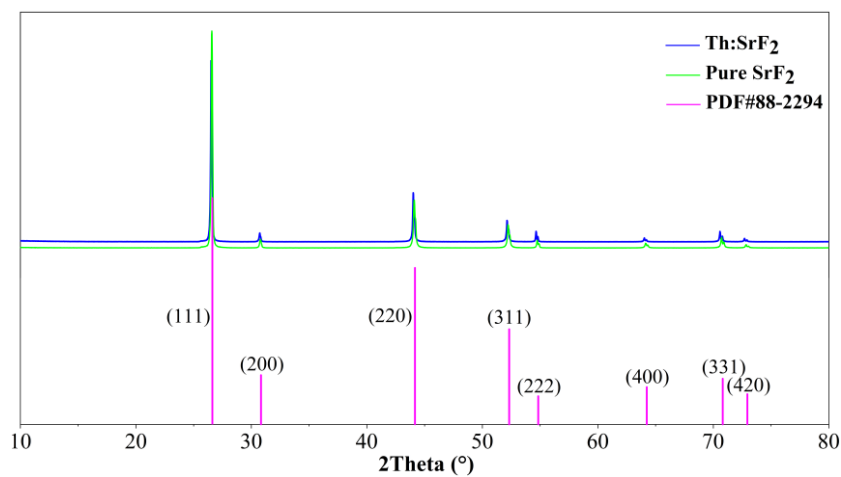

**Figure 4.** The XRD patterns of Th:SrF$_2$ crystals and pure SrF$_2$ crystal.

**Figure 4** shows the XRD patterns of as-grown pure SrF$_2$ and Th:SrF$_2$ crystals. As can be seen that the diffraction peaks of the as-grown crystals correspond well to the standard Powder Diffraction File (PDF#88-2294) of SrF$_2$. It is noteworthy that, compared to the pure SrF$_2$ crystal, the positions of the diffraction peaks of Th:SrF$_2$ crystal shift slightly to the left, indicating that Th doping causes an increase in cell parameters and unit cell volume. We further fitted the cell parameters and unit cell volume of the as-grown crystals based on XRD diffraction data and calculated the structural changes caused by Th doping. In the theoretical research section, we have determined the optimal defect configuration, namely the Th$_{Sr}$-2F$_{in}$ (180°) defect model. Due to the fact that the model was constructed in the SrF$_2$ supercell of 2×2×2, with the corresponding chemical formula Th$_1$Sr$_{31}$F$_{66}$, the weight percentage of Th in the theoretical model is approximately 5.57 wt.%, which is at a similar level to the Th:SrF$_2$ crystals we experimentally grown. The experimental and theoretical comparison of structural changes in Th:SrF$_2$ crystals relative to pure SrF$_2$ crystals are listed in **Table 3**. As can be seen that, in terms of structural changes in crystals, there is a high degree of consistency between the experimental and theoretical results (especially the Th$_{Sr}$-2F$_{in}$ (180°) model). This to some extent supports the judgment of our theoretical research on the optimal defect configuration.

**Table 3.** Experimental and theoretical comparison of structural changes in Th: SrF2 crystals relative to pure SrF$_2$ crystals.

| Structural changes | Δa [%] | ΔV [%] | Weight percentage of Th [wt. %] |
|---|---|---|---|
| Experiment (as-grown Th:SrF$_2$) | 0.117~0.168 | 0.353~0.507 | 5.72~6.06 |
| Theory (Th$_{Sr}$-2F$_{in}$ (180°) model) | 0.116 | 0.347 | 5.57 |
| Theory (Th$_{Sr}$-2F$_{in}$ (90°) model) | 0.261 | 0.783 | 5.57 |

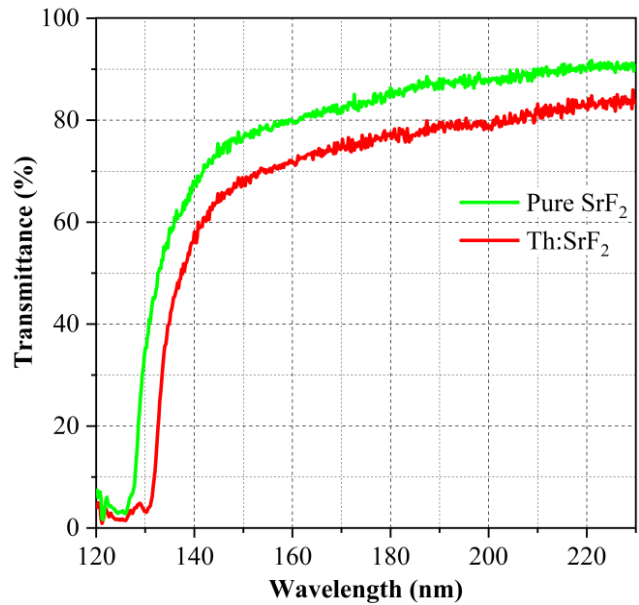

**Figure 5**. The VUV transmittance curves (120-230 nm) of the as-grown pure SrF$_2$ and Th:SrF$_2$ crystal samples with a thickness of 1 mm.

**Figure 5** present the vacuum ultraviolet (VUV) transmittance of the as grown pure $SrF_2$ and $Th:SrF_2$ crystal samples with a thickness of 1 mm. As can be seen that, compared to the pure $SrF_2$ crystals, the UV cutoff edge of $Th:SrF_2$ crystal does indeed undergo a slight redshift, which is consistent with our theoretical results. In addition, the transmittance of the $Th:SrF_2$ crystal is as high as ~69% at 149 nm, which is only slightly lower than that of the pure $SrF_2$ crystal (~77% at 149 nm). The result is exciting, at such high doping levels (>$6\times10^{20}$ cm$^{-3}$), $Th:SrF_2$ crystals still maintain a considerable high level of VUV transmittance at the nuclear transition band. This reflects the excellent tolerance of $SrF_2$ crystal for Th ions, that even with such high concentrations of doping, its transmittance has not significantly deteriorated. This is undoubtedly another advantage of $Th:SrF_2$ crystal as a candidate for solid-state nuclear clocks material. Taking into account the unique advantages of $Th:SrF_2$ crystal in Th doping and VUV transmittance, we consider that $Th:SrF_2$ crystal will become the preferred choices for the future research on solid-state nuclear spectroscopy and solid-state nuclear clock.

Finally, considering the future application prospects of $Th:SrF_2$ crystal in solid-state nuclear spectroscopy and nuclear clocks, we would like to briefly explore the potential of $Th:SrF_2$ crystals in background luminescence and radiation damage resistance, based on existing literature results. Firstly, due to the similar refractive indices of $SrF_2$ (1.59) and $CaF_2$ (1.58) at 150 nm [59], the background luminescence caused by Cherenkov radiation in the nuclear transition band is expected to be acceptable. Secondly, the self-trapped exciton emission peak of $SrF_2$ crystal is located at ~300 nm, and the background emission in the nuclear transition band is almost zero [60,61]. After Th doping, it can be predicted that the self-limiting exciton emission peak will further redshift due to the reduction of the band gap, this means that the background fluorescence of $Th:SrF_2$ crystal will not cause obvious interference to the nuclear transition signal.

In addition, as reviewed in the introduction, the radiation damage caused by the strong α radioactivity ($^{229}$Th decays 100% via an *α*-decay of 5.167 MeV) of $^{229}$Th will color the $Th:CaF_2$ crystals and cause a sharp decrease in its transmittance in the nuclear transition band [31]. What we are actually concerned about is whether $SrF_2$, like $CaF_2$, will be colored and exhibit poor VUV transmittance due to the radiation damage and defects induced by $^{229}$Th at a relatively high doping concentration. Research has shown that $SrF_2$ crystals were not colored like $CaF_2$ crystal under strong alpha irradiation from a $^{238}$PuO$_2$ source (the cumulative dose of $3\times10^{20}$ alpha particles/m$^2$, with energy of 5.5-5.0 MeV/particle) [62]. It seems to suggest another advantage of $Th:SrF_2$ crystal as a candidate for solid-state nuclear clocks material, although the actual situation still needs to be further confirmed in future experimental research on $^{229}$Th:SrF$_2$ crystal.

In summary, this article discloses a highly promising candidate for solid-state nuclear optical clock material, Th:SrF$_2$ crystal. The segregation coefficient for thorium in this crystal is close to 1, which can ensure highly efficient, uniform and low loss doping of Th. This means the difficulty of cultivating solid-state nuclear clock materials will be greatly reduced. The reason why it has such doping characteristics is investigated by the theoretical calculations from the perspective of the equilibrium of thermodynamic and kinetic factors. The cultivated $^{232}$Th:SrF$_2$ crystal exhibits excellent VUV transmittance in nuclear transition band (~69% at 149 nm) while meeting an extremely high doping level (>6×10$^{20}$ cm$^{-3}$), reflecting excellent tolerance for Th nuclei. Moreover, the crystal is expected to have certain advantages in radiation damage resistance and background luminescent noise, it will not be colored under strong alpha irradiation and can fully unleash its high concentration doping characteristics. The authors think that the discovery of Th:SrF$_2$ crystal is of great importance for the development of solid-state nuclear clock materials and it will become the preferred material for the solid-state nuclear clock research in the future.

We acknowledge support by Zhangjiang laboratory (ZJSP21A001D) and National Natural Science Foundation of China (NSFC) (12341042, 12341403).